\documentclass[showpacs,aps,prd,twocolumn]{revtex4}
\usepackage{amsmath}
\usepackage{epsfig}
\usepackage{subfigure}

\newcommand{\ep}{\varepsilon}

\newcommand{\eqs}[1]{\begin{equation} \begin{split} #1\end{split} \end{equation} }

\newcommand{\ga}{\gamma^5}
\newcommand{\gmu}{\gamma^\mu}

\newcommand{\gsig}{\gamma^\sigma}

\newcommand{\ie}{{\it i.e.}}
\newcommand{\eg}{{\it e.g.}}
\newcommand{\etal}{{\it et al.}}

\newcommand{\ce}[1]{Eq.~(\ref{#1})}

\newcommand{\ct}[1]{{Table~(\ref{#1})}}

\begin{document}

\bigskip
\title{Two-photon width of $\eta_c$ and $\eta_c'$ from Heavy-Quark Spin Symmetry}

\author
{J.P. Lansberg$^{a,b}$ and T.N. Pham$^{a}$}

\affiliation{
$^{a}$Centre de Physique Th\'eorique, \\
Centre National de la Recherche Scientifique, UMR 7644,\\ {\'E}cole
Polytechnique, 91128 Palaiseau, France \\
\\$^{b}$Physique Th\'eorique Fondamentale, Universit\'e de Li\`ege,\\
17 All\'ee du 6 Ao\^ut, B\^at. B5, B-4000 Li\`ege-1, Belgium}

 
\begin{abstract}
\parbox{14cm}{\rm 

We evaluate the two-photon width of the pseudoscalar charmonia, $\Gamma_{\gamma \gamma}(\eta_c)$
and $\Gamma_{\gamma \gamma}(\eta'_c)$, within a Heavy-Quark Spin-Symmetry setting and show 
that whereas the former width agrees with experiment, the latter  is more than 
twice larger than the recent measurement by CLEO. When binding-energy effects are 
included in the $\eta'_c$ case, the discrepancy is worse, pointing out at a possible anomaly in 
the $\eta'_c$ decay.}
\end{abstract}
\pacs{13.20.Gd 13.25.Gv 11.10.St 12.39.Hg}
\maketitle

\section{Introduction}

Whereas heavy-quarkonium production is still a great source of debates 
(see \cite{yr,Lansberg:2006dh} for recent reviews), the physics of quarkonium 
decay seems to be better understood within the conventional framework of QCD.
However, a recent estimation of the ratio of the two-photon width of the $\eta'_c$ 
to that of the $\eta_c$  by the CLEO collaboration~\cite{Asner:2003wv} 
seems to contradict most of the theoretical 
predictions~\cite{Ackleh:1991dy,Kim:2004rz,Ahmady:1994qf}. 
Indeed, by assuming ${\cal B} (\eta_c \to KK \pi)={\cal B} (\eta'_c \to KK \pi)$,
they have obtained $\Gamma_{\gamma \gamma}(\eta'_c)= 1.3 \pm 0.6$ keV, whereas the predictions
of~\cite{Ackleh:1991dy,Kim:2004rz,Ahmady:1994qf} range from $3.7$ to $5.7$ keV.

It is our purpose here to have another look at this problem using an effective
Lagrangian procedure satisfying heavy-quark spin symmetry and including binding energy or
equally mass effects, which would take into account features typical of
radially-excited states. 

Indeed, the $\eta'_c$ is the first radially-excited pseudoscalar charmonium,
labeled in the spectroscopic notation by $2 ^1S_0$, with a mass $M_{\eta'_c}=3638\pm 5$ 
MeV~\cite{Eidelman:2004wy}, 
that is noticeably higher than for the ground states $\eta_c$ and $J/\psi$. Its first observation was
done by the Belle collaboration~\cite{Choi:2002na} in $B\to K K_S K^- \pi^+$ decay and
was further confirmed by {\sc BaBar}~\cite{Aubert:2003pt}.

As a consequence of $C$-conservation, $\eta'_c$, like $\eta_c$,
can decay into two photons, which is from a theoretical point of
 view a rather clean channel to analyze. There have been several 
calculations of the $\eta'_c\to \gamma\gamma$ in the literature, 
some following Bethe-Salpeter equation~\cite{Chao:1996bj}, following 
Salpeter equation or relativistic quark models~\cite{Kim:2004rz,Munz:1996hb,Ebert:2003mu,Ackleh:1991dy}, 
 and some based on the 
non-relativistic results (see for instance~\cite{Kwong:1987mj})
but taking into account differences in the singlet and triplet wave function
at the origin \cite{Ahmady:1994qf,Gerasimov:2005pk}. In particular, 
non-relativistic calculations can only be done by considering the
$\eta'_c$  as a 2$S$ state with
the same mass,  $2m_{c}$, as the 1$S$ state with the result  that the
calculated decay rate differs from the $\eta_c$ one
only through the wave function at the origin, see Eq (3.17) 
of~\cite{Kwong:1987mj}, or the long distance
matrix element of NRQCD, see Eqs. (4.17) and (4.19) of~\cite{yr}.

Since the $\eta'_c$ is more than $600\rm\,MeV$ above the
$\eta_c$, the mass effects on the decay rate could be important. A better
approach, which would allow the inclusion of such effects, 
would be to use  relativistic kinematics
in the calculation of the width. For this purpose, we need 
to construct an effective Lagrangian for the process 
$c\bar{c}\to \gamma\gamma$ by expanding the charm-quark propagator
in powers of $q^{2}/m_{c}^{2}$ , with $q=p_{c} -p_{\bar{c}}$,
and  neglecting terms of ${\cal O}(q^{2}/m_{c}^{2})$ terms. The 
propagator will now depend only on the charm-quark mass and the 
binding energy of the charmonium state~\cite{Kuhn:1979bb,Pham}.

The effective Lagrangian derived in our approach
will then allow a calculation of the decay amplitude in terms of the
 matrix element of a local operator. The latter is, for 
the two-photon decay width of $\eta_{c}$
and $\eta_{c}^{\prime}$ , the matrix element for the axial-vector current
$\bar{c}\gamma_{\mu}\gamma_{5}c$
between the vacuum and $\eta_{c}$ or $\eta_{c}^{\prime}$.

The non-perturbative parameters are here the decay constant 
$f_{n^1S_0}$ and $f_{n^3S_1}$ which can be  given
by the spatial wave function at the origin $\psi(0)$~\cite{Novikov:1977dq}. 
It should be stressed here that our approach differs from the traditional approach in an
important way. We express the decay amplitude in terms of the matrix 
element of a local operator which could be measured or extracted 
from measured physical quantities, like the leptonic-decay constant
 or could also be computed via sum rules~\cite{Reinders:1982zg} or lattice 
simulations~\cite{Dudek:2006ej}.

We shall rely on the heavy-quark spin-symmetry (HQSS) 
relations~\cite{neubert,nardulli,DeFazio}
which state the equality between $f_{\eta_c}$ and $f_{J/\psi}$
and between $f_{\eta^{\prime}_c}$ and $f_{\psi^{\prime}}$. The derivation
of these relations is based on the fact that, in our approach,  the 
flavor-conserving charm-quark
currents $\bar{c}\,\gamma_{\mu}\,c$ and $\bar{c}\,\gamma_{\mu}\gamma_{5}\,c$
take the form of  an effective current in which 
 $c$ and $\bar{c}$ are replaced by static heavy quark field operator
and the ${\cal O}(1/(2 m_{c}))$ terms are neglected.

In this paper, we shall first derive the effective Lagrangian
for the decay of singlet 1$S$ state of charmonium into two photons.
We then show that this effective Lagrangian, combined with HQSS, gives the same result as the
traditional non-relativistic approach and produces a decay rate for $\eta_{c}$
in agreement with measurement. In the next section, we use this Lagrangian
to determine the $\eta^{\prime}_{c}$ two-photon decay rate in terms of the
$\psi^{\prime}$ leptonic width, as our main purpose here is to see whether 
our approach, which works for the $\eta_{c}$, could explain the 
observed decay rate of the $\eta^{\prime}_{c}$ into 
two photons.

\section{Effective Lagrangian for $^1S_0$ decay into two photons}

As announced, we now write down an effective Lagrangian for the coupling of the $c \bar c $ pair to 
two photons and to a dilepton pair $\ell \bar \ell$:
\eqs{
{\cal L}^{\gamma \gamma}_{\rm eff}=&-i c_1(\bar c \gsig \ga c) \ep_{\mu \nu \rho \sigma} F^{\mu\nu} A^\rho\\
{\cal L}^{\ell \bar \ell}_{\rm eff}=&-  c_2(\bar c \gmu c) (\ell \gamma_\mu \bar \ell)
}
with $\displaystyle c_1\simeq \frac{Q_c^2 (4\pi
  \alpha_{em})}{M_{\eta_c}^2+b_{\eta_c} M_{\eta_c}}$ 
 and 
$\displaystyle c_2=\frac{Q_c (4\pi \alpha_{em})}{M_\psi^2}$. 

The factor $1/(M_{\eta_c}^2+b_{\eta_c} M_{\eta_c})$ in $c_{1}$ 
contains the binding-energy effect~\cite{Kuhn:1979bb,Pham} 
and is obtained from the denominator of the charm-quark 
propagator ($k_{1}, k_{2}$  being the outgoing-photon momenta):
\begin{equation}
\frac{1}{[(k_{1}-k_{2})^{2}/4 - m_{c}^{2}]}
\end{equation}
by neglecting the term containing the relative momenta $q=p_c-p_{\bar c}$ of the quarks.
For real photons, this factor can be written as
\begin{equation}
-\frac{1}{[ (M^{2} +bM)/2]}
\end{equation}
with $b$ $(=2m_{c} -M)$, the bound-state binding energy
and $M$ the charmonium mass (in order to be consistent, 
we keep only term linear in $b$, since the $O(q^{2}/m_{c}^{2})$ terms
have been neglected in the propagator). 

\begin{figure}[h]
\centering
\subfigure[~$cc\to \gamma \gamma$]{\includegraphics[height=2.5cm,angle=0]{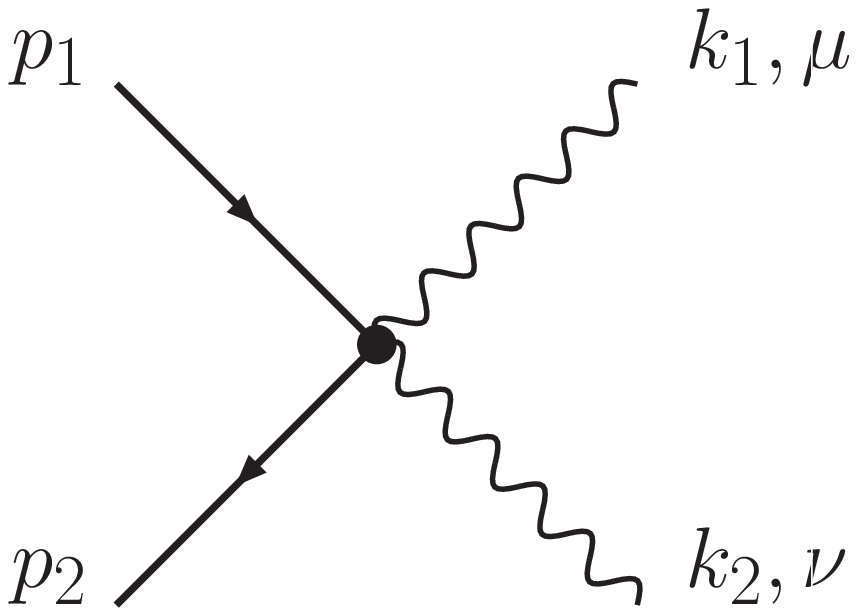}}\quad \quad
\subfigure[~$cc\to \ell \bar \ell$]{\includegraphics[height=2.5cm,angle=0]{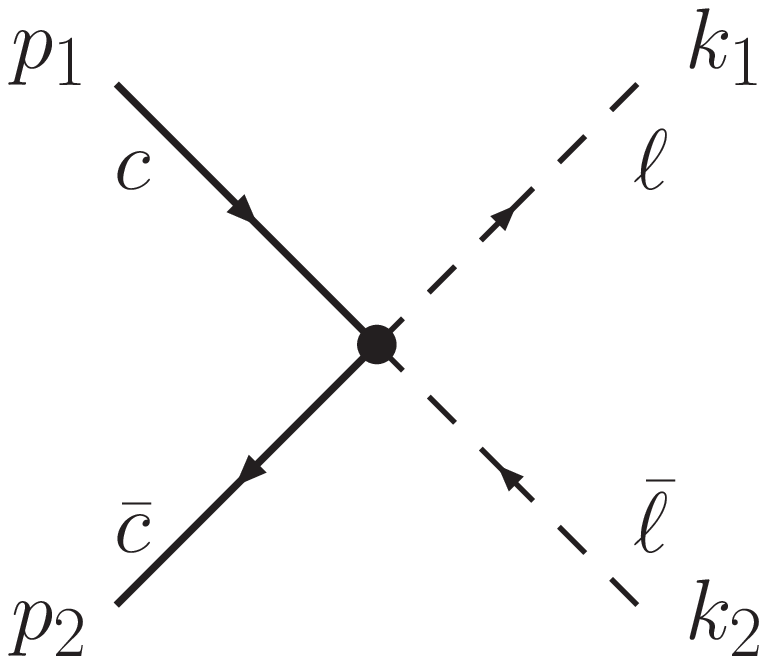}}
\caption{Effective coupling between a $c \bar c$ pair and two photons (a) and
a dilepton pair (b).}
\label{fig1}
\end{figure}

\section{Heavy-quark spin-symmetry prediction for $\Gamma_{\gamma \gamma}(\eta_c)$}

First, we want to redo the calculation of $\Gamma_{\ell \bar \ell}(\psi)$
and  $\Gamma_{\gamma \gamma}(\eta_c)$ through the simple 
application of heavy-quark spin symmetry (HQSS) and to show that 
the results are identical to those of non-relativistic calculations.

Defining $ \left<0|\bar c \gmu  c| \psi \right>\equiv f_{\psi}
M_\psi\ep^\mu$, we have the following expression for the amplitude 
for $\psi \to \ell \bar \ell$:
\eqs{
{\cal M}_{\ell \bar \ell}= 
Q_c (4 \pi \alpha_{em})\frac{f_{\psi}}{M_\psi} \ep^\mu (\ell \gamma_\mu \bar \ell)
}
from which we obtain the width (neglecting the lepton masses):
\eqs{
\Gamma_{\ell \bar \ell}(\psi)=\frac{1}{64 \pi^2 M_\psi} \int d \Omega |{\cal M}|^2=\frac{4 \pi Q_c^2 \alpha^2_{em} f_\psi^2}{3 M_\psi}.
}
Using $M_{\psi}f_\psi^2=12 |\psi(0)|^2$~\cite{Novikov:1977dq}, we recover the non-relativistic result of
Kwong~\etal~\cite{Kwong:1987mj}.
The experimental value for the leptonic width of the $J/\psi$ 
($\Gamma_{e^+ e^-}(J/\psi)=5.40 \pm 0.15 \pm 0.07$ keV~\cite{Eidelman:2004wy}) and its mass (3.097 GeV)
fixes --omitting NLO corrections for now-- $f_{J/\psi}$ at $410$ MeV. For the $\psi'$, we
correspondingly get $f_{\psi'}$ at $279$ MeV for $\Gamma_{e^+ e^-}(\psi')=2.10 \pm 0.12$ keV~\cite{Eidelman:2004wy} and 
$M_{\psi'}=3.686$ GeV.

Similarly,  with $ \left<0|\bar c \gmu \ga c| \eta_c \right>\equiv i f_{\eta_c} P^\mu$, the amplitude for
$\eta_c \to \gamma \gamma$ is readily obtained:
\eqs{
{\cal M}_{\gamma \gamma}=- 4 i Q_c^2 (4 \pi \alpha_{em})\frac{f_{\eta_c}}{M^2_{\eta_c}+b_{\eta_c} M_{\eta_c}} \epsilon_{\mu \nu \rho \sigma} 
\ep_1^\mu \ep_2^\nu k_1^\rho  k_2^\sigma
}
from which we obtain the $\eta_c(1S)$ width (with $b_{\eta_c}\simeq 0$):
\eqs{
\Gamma_{\gamma \gamma}(\eta_c)=\frac{1}{2}\frac{1}{64 \pi^2 M_{\eta_c}} \int d \Omega |{\cal M}|^2=\frac{4 \pi Q_c^4 \alpha^2_{em} f_{\eta_c}^2}{M_{\eta_c}},
}
the factor $\frac{1}{2}$ being the Bose-symmetry factor.

As suggested by HQSS, let us now suppose the equality between $f_{J/\psi}$ and
$f_{\eta_c}$, enabling us the following  evaluation, 
$\Gamma_{\gamma \gamma}(\eta_c)=7.46$ keV.

When NLO corrections are taken into account~\cite{Kwong:1987mj},
\eqs{\Gamma^{NLO}(^3S_1)&= \Gamma^{LO} \left(1- \frac{\alpha_s}{\pi}\frac{16}{3}\right) \\
\Gamma^{NLO}(^1S_0)&= \Gamma^{LO} \left(1- \frac{\alpha_s}{\pi}\frac{(20-\pi^2)}{3}\right),}
with $\alpha_s=0.26$, $\Gamma_{\gamma \gamma}(\eta_c)$ is shifted to 9.66 keV.
The latter agrees with the world average value
$7.4 \pm 0.9 \pm 2.1$ keV~\cite{Eidelman:2004wy} in view of the large statistic and systematic 
uncertainties in the measured value. This indicates that our effective Lagrangian approach can also
successfully predict the $\eta_c$ two-photon width. The agreement with experiment also suggests
that there is no large spin-symmetry breaking term in the charm vector and axial-vector current
matrix elements. We now use the same effective Lagrangian and HQSS to compute the $\eta'_c$
two-photon width.

\section{HQSS predictions for $\Gamma_{\gamma \gamma}(\eta'_c)$}

We now turn to the excited states. 
Extrapolating HQSS to $2S$ states,~\ie~$f_{\psi'}=f_{\eta'_c}$, and neglecting
binding energy effects, we obtain
$\Gamma^{}_{\gamma \gamma}(\eta'_c)= \Gamma^{}_{\gamma \gamma}(\eta_c) 
\frac{f^2_{\psi'}}{f^2_{J/\psi}}=3.45~\hbox{keV}$,
which is more than twice larger than the evaluation by CLEO ($1.3\pm 0.6$ keV) 
although nearly in agreement with 
Ackleh~\etal~\cite{Ackleh:1991dy} ($3.7$ keV), Kim~\etal~\cite{Kim:2004rz} ($4.44\pm 0.48$ keV), 
Ahmady~\etal~\cite{Ahmady:1994qf} ($5.7\pm 0.5\pm0.6$ keV).

Binding-energy effects are easily taken into account by introducing a correcting factor such that 
$\Gamma_{\gamma \gamma}(\eta'_c)$ can be written as
as a function of $\Gamma_{\gamma \gamma}(\eta_c)$, $\Gamma_{e^+e^-}(J/\psi)$ and $\Gamma_{e^+e^-}(\psi')$
as follows
\eqs{\label{eq:g_etacp}
\Gamma_{\gamma \gamma}(\eta'_c)
=\Gamma_{\gamma \gamma}(\eta_c)
\left(\left(\frac{M_{\eta_c}^2+b_{\eta_c} M_{\eta_c}}{M_{\eta'_c}^2+b_{\eta'_c} M_{\eta'_c}}\right)^2 
 \frac{M_{\eta'_c}^3}{M_{\eta_c}^3}\right) \\
\times \left(\frac{\Gamma_{e^+e^-}(\psi')}{\Gamma_{e^+e^-}(J/\psi)}
  \frac{M_{\psi'}}{M_{J/\psi}}\right).}

This gives 
\eqs{\Gamma^{}_{\gamma \gamma}(\eta'_c)=4.1~\hbox{keV},}
therefore the introduction of differences in the mass of $\eta_c$ and $\eta'_c$ increases 
the discrepancy with the experimental result obtained by CLEO.
Note that, up to corrections due to differences in the scale of $\alpha_s$, the radiative corrections 
 are canceled in \ce{eq:g_etacp} as well as in the formula giving the first quoted value, 3.45 keV.
If one wanted to introduce relativistic corrections in the spirit of NRQCD, one would expect
them to cancel also, following Eqs (4.3c), (4.3d), (A31), (A32c), (A34) and (A35a) of 
Bodwin~\etal~\cite{Bodwin:1994jh}.

It has to be noted however that the experimental values of $\Gamma_{\gamma \gamma}(\eta_c)$
are affected by a large systematic uncertainty related to the branching
${\cal B} (\eta_c \to KK \pi)$ and the evaluation of  $\Gamma_{\gamma \gamma}(\eta'_c)$
done by CLEO was realised by assuming ${\cal B} (\eta_c \to KK
\pi)={\cal B} (\eta'_c \to KK \pi)$ which is only to hold approximately. 
This assumption also allows an extraction of ${\cal B}(B \to K\eta'_{c})$
from the Belle measurement of the ratio 
$({\cal B}(B \to K\eta'_{c})\times {\cal B} (\eta'_c \to KK\pi))/({\cal B}(B \to
K\eta_{c})\times {\cal B} (\eta_c \to KK\pi)$~\cite{Choi:2002na}. The value 
of the ratio
${\cal B}(B \to K\eta'_{c})/{\cal B}(B \to K\eta_{c})$ thus obtained 
seems to agree with a theoretical
prediction using QCD factorisation model for  colour-suppressed $B$ decays 
with a charmonium in the final state~\cite{Song}. Thus the assumption of
the approximate  equality between the $ \eta'_c \to KK\pi$ and 
$ \eta_c \to KK\pi$ branching ratio seems to be justified to some extent. In this 
case, the CLEO low value for  the ratio $({\cal B}(\eta'_{c}\to \gamma
\gamma){\cal B} (\eta'_c \to KK\pi))/({\cal B}(\eta_{c}\to \gamma \gamma){\cal B}
(\eta_c \to KK\pi))$ would imply the small $\eta'_{c}\to \gamma\gamma$
decay rate quoted above.

There exist however models that are able to reproduce correctly  $\Gamma_{\gamma \gamma}(\eta'_c)$ 
but, in general, they  tend to underestimate $\Gamma_{\gamma \gamma}(\eta_c)$.  Indeed, 
M\"unz~\cite{Munz:1996hb} predicts $3.5\pm 0.4$ keV for
$\eta_c$  and $1.38 \pm 0.3$ keV for $\eta'_c$, Chao~\etal~\cite{Chao:1996bj} $5.5$ keV for
$\eta_c$  and $2.1$ keV for $\eta'_c$, and Ebert~\etal~\cite{Ebert:2003mu} $5.5$ keV for
$\eta_c$  and $1.8$ keV for $\eta'_c$ (see also the results of~\cite{Crater:2006ha}). This 
clearly points at a specificity not yet understood of the $\eta'_c$ decay.
All the theoretical predictions and the experimental measurements can be found in \ct{tab-res}.

\begin{widetext}

\begin{table}[h]
\begin{tabular}{|c|c|c|c|c|c|c|c|c|}
\hline
$\Gamma_{\gamma \gamma}$ & Experiments & This paper &Ackleh~\cite{Ackleh:1991dy}& Kim~\cite{Kim:2004rz}&
Ahmady~\cite{Ahmady:1994qf} &
M\"unz~\cite{Munz:1996hb} &Chao~\cite{Chao:1996bj}&Ebert~\cite{Ebert:2003mu}\\
\hline\hline
$\eta_c$ & $7.4 \pm 0.9 \pm 2.1$ (PDG~\cite{Eidelman:2004wy}) & $7.5-10$ & $4.8$ & $7.14 \pm 0.95$&$11.8\pm0.8\pm 0.6$&
$3.5\pm 0.4$&$5.5$&$5.5$ 
\\
$\eta'_c$& $1.3 \pm 0.6$ (CLEO~\cite{Asner:2003wv})& $3.5-4.5$ & $3.7$&$4.44\pm 0.48$&$5.7\pm 0.5\pm0.6$&
$1.38 \pm 0.3$
&$2.1$&$1.8$ \\
\hline
\end{tabular}
\caption{Summary of experimental measurements and theoretical predictions for 
$\Gamma_{\gamma \gamma}(\eta_c)$ and $\Gamma_{\gamma \gamma}(\eta'_c)$. (All values
are in units of keV).}\label{tab-res}
\end{table}

\end{widetext}

\section{Conclusion}

Whereas heavy quarkonia are supposed to be reasonably described by non-relativistic approximations, 
some works (\eg~\cite{Chiang:1994qi,Lansberg:2005pc,Lansberg:2005aw,smith}) have pointed out that
non-static effects within quarkonium (especially in radially-excited states)  
should not be neglected without further considerations.

In the case  of $\eta_c$, we have seen that the simple application of HQSS gives a reasonable estimate 
of the width compared to  the world-average experimental measurements. For $\eta'_c$,
we have obtained the same discrepancy as other models.
On the other hand, we have shown here that the introduction of binding energy 
in this calculation introduces a correction of about 20 \% but worsens the 
comparison with the CLEO measurement.

When one considers the ratio of the two decay widths, radiative corrections 
cancel out, up to effects due to changes in the renormalisation scale. This
might slightly affect the results, but not sufficiently to recover agreement
with data. Of course, heavy-quark spin symmetry (or, equivalently, the equality between 
the decay constant  for the $^3S_1$ and the $^1S_0$) could be broken 
for excited states, but it is quite unlikely that it could be so badly broken to
 explain such a discrepancy.

Since many other works have shown difficulties to reproduce both $\eta_c$ and $\eta'_c$ two-photon widths,
 we are looking forward for a confirmation of the CLEO measurements, especially through a better
understanding of their branching ratio in $K K \pi$.


\begin{thebibliography}{99}

\bibitem{yr}
N.~Brambilla {\it et al.}, {\it Heavy quarkonium physics}, CERN Yellow Report, CERN-2005-005, 
2005  Geneva : CERN, 487 pp 
[arXiv:hep-ph/0412158].

\bibitem{Lansberg:2006dh}
  J.~P.~Lansberg,
  {\it $J/\psi$, $\psi'$ and $\Upsilon$ production at hadron colliders: A review},
  arXiv:hep-ph/0602091.

\bibitem{Asner:2003wv}
  D.~M.~Asner {\it et al.}  [CLEO Collaboration],
  Phys.\ Rev.\ Lett.\  {\bf 92} (2004) 142001
  [arXiv:hep-ex/0312058].


\bibitem{Ackleh:1991dy}
  E.~S.~Ackleh and T.~Barnes,
  Phys.\ Rev.\ D {\bf 45} (1992) 232.


\bibitem{Kim:2004rz}
  C.~S.~Kim, T.~Lee and G.~L.~Wang,
  Phys.\ Lett.\ B {\bf 606} (2005) 323
  [arXiv:hep-ph/0411075].

\bibitem{Ahmady:1994qf}
  M.~R.~Ahmady and R.~R.~Mendel,
  Phys.\ Rev.\ D {\bf 51} (1995) 141
  [arXiv:hep-ph/9401315].


\bibitem{Eidelman:2004wy}
  S.~Eidelman {\it et al.}  [Particle Data Group],
  Phys.\ Lett.\ B {\bf 592} (2004) 1.



\bibitem{Choi:2002na}
  S.~K.~Choi {\it et al.}  [BELLE collaboration],
  Phys.\ Rev.\ Lett.\  {\bf 89} (2002) 102001
  [Erratum-ibid.\  {\bf 89} (2002) 129901]
  [arXiv:hep-ex/0206002].

\bibitem{Aubert:2003pt}
  B.~Aubert {\it et al.}  [BABAR Collaboration],
  Phys.\ Rev.\ Lett.\  {\bf 92} (2004) 142002
  [arXiv:hep-ex/0311038].

\bibitem{Chao:1996bj}
  K.~T.~Chao, H.~W.~Huang, J.~H.~Liu and J.~Tang,
  Phys.\ Rev.\ D {\bf 56}, 368 (1997)
  [arXiv:hep-ph/9601381].


\bibitem{Munz:1996hb}
  C.~R.~Munz,
  Nucl.\ Phys.\ A {\bf 609}, 364 (1996)
  [arXiv:hep-ph/9601206].

\bibitem{Ebert:2003mu}
  D.~Ebert, R.~N.~Faustov and V.~O.~Galkin,
  Mod.\ Phys.\ Lett.\ A {\bf 18}, 601 (2003)
  [arXiv:hep-ph/0302044].


\bibitem{Kwong:1987mj}
  W.~Kwong, J.~L.~Rosner and C.~Quigg,
  Ann.\ Rev.\ Nucl.\ Part.\ Sci.\  {\bf 37} (1987) 325.




\bibitem{Gerasimov:2005pk}
  S.~B.~Gerasimov and M.~Majewski,
  arXiv:hep-ph/0504067.





\bibitem{neubert}
 M.~Neubert,
Phys.\ Rept.\  {\bf 245}, 259 (1994);
[arXiv:hep-ph/9306320].


\bibitem{DeFazio}
  F.~De Fazio, in {\it At the Frontier of Particle Physics/Handbook
of QCD}, edited by M. A. Shifman (World Scientific, 2001) 1671
[arXiv:hep-ph/0010007]~;
  arXiv:hep-ph/0010007.


\bibitem{nardulli}
  R.~Casalbuoni, A.~Deandrea, N.~Di Bartolomeo, R.~Gatto, F.~Feruglio and G.~Nardulli,
  Phys.\ Rept.\  {\bf 281}, 145 (1997)
  [arXiv:hep-ph/9605342].





\bibitem{Kuhn:1979bb}
  J.~H.~Kuhn, J.~Kaplan and E.~G.~O.~Safiani,
  Nucl.\ Phys.\ B {\bf 157} (1979) 125.


\bibitem{Pham}
  T.~N.~Pham and G.~h.~Zhu,
  Phys.\ Lett.\ B {\bf 619} (2005) 313
  [arXiv:hep-ph/0412428].

\bibitem{Novikov:1977dq}
  V.~A.~Novikov, L.~B.~Okun, M.~A.~Shifman, A.~I.~Vainshtein, 
  M.~B.~Voloshin and V.~I.~Zakharov,
  Phys.\ Rept.\  {\bf 41} (1978) 1.

\bibitem{Reinders:1982zg}
  L.~J.~Reinders, H.~R.~Rubinstein and S.~Yazaki,
  Phys.\ Lett.\ B {\bf 113} (1982) 411.


\bibitem{Dudek:2006ej}
  J.~J.~Dudek, R.~G.~Edwards and D.~G.~Richards,
  Phys.\ Rev.\ D {\bf 73} (2006) 074507
  [arXiv:hep-ph/0601137].

\bibitem{Song}
  Z.~z.~Song, C.~Meng and K.~T.~Chao,
  Eur.\ Phys.\ J.\ C {\bf 36}, 365 (2004)
  [arXiv:hep-ph/0209257].



\bibitem{Bodwin:1994jh}
  G.~T.~Bodwin, E.~Braaten and G.~P.~Lepage,
  Phys.\ Rev.\ D {\bf 51}, 1125 (1995)
  [Erratum-ibid.\ D {\bf 55}, 5853 (1997)]
  [arXiv:hep-ph/9407339].


\bibitem{Crater:2006ha}
  H.~W.~Crater, C.~Y.~Wong and P.~Van Alstine,
  arXiv:hep-ph/0603126.


\bibitem{Chiang:1994qi}
  H.~C.~Chiang, J.~Hufner and H.~J.~Pirner,
  Phys.\ Lett.\ B {\bf 324}, 482 (1994)
  [arXiv:hep-ph/9401233].



\bibitem{Lansberg:2005pc}
  J.~P.~Lansberg, J.~R.~Cudell and Y.~L.~Kalinovsky,
  Phys.\ Lett.\ B {\bf 633} (2006) 301
  [arXiv:hep-ph/0507060].



\bibitem{Lansberg:2005aw}
J.~P.~Lansberg, {\it Quarkonium Production at High-Energy Hadron Colliders}, Ph.D. 
Thesis, ULg, Li\`ege, Belgium, 2005 [arXiv:hep-ph/0507175].


\bibitem{smith}
C. Smith, {\it Bound State Description in Quantum Electrodynamics and Chromodynamics }, Ph.D. 
Thesis, UCL, Louvain-la-Neuve, Belgium, 2002.



\end{thebibliography}
\end{document}